\documentclass[aps,prl,showpacs,twocolumn]{revtex4-1}
\usepackage[latin1]{inputenc}
\usepackage[T1]{fontenc}
\usepackage[french,dutch,english]{babel}
\usepackage{amsmath}
\usepackage{amssymb,amsfonts,textcomp}
\usepackage{color}
\usepackage{array}
\usepackage{hhline}
\usepackage{graphicx}
\usepackage{hyperref}
\usepackage{pdfpages}
\hypersetup{pdftex, colorlinks=true, linkcolor=blue, citecolor=blue, filecolor=blue, urlcolor=blue, pdftitle=Majorana paper, pdfauthor=Stevan Nadj-Perge, pdfsubject=, pdfkeywords=}

\makeatletter
\newcommand\ps@Standard{
  \renewcommand\@oddhead{}
  \renewcommand\@evenhead{}
  \renewcommand\@oddfoot{}
  \renewcommand\@evenfoot{\@oddfoot}
  \renewcommand\thepage{\arabic{page}}
}
\makeatother

\begin{document}
\title{Majorana fermions in chains of magnetic atoms on a superconductor}

\author{S. Nadj-Perge, I. K. Drozdov, B. A. Bernevig, and Ali Yazdani }
\affiliation{Joseph Henry Laboratories and Department of Physics, Princeton University, Princeton, New Jersey 08544}
\date{\today}
\begin{abstract}
We propose an easy-to-build easy-to-detect scheme for realizing Majorana fermions at the ends of  a chain of magnetic atoms on the surface of a superconductor. Model calculations  show that such chains can be easily tuned between trivial and topological ground state. In the latter, 
spatial resolved spectroscopy can be used to probe the Majorana fermion end states. Decoupled 
Majorana bound states can form even in short magnetic chains consisting of only tens of atoms. We propose scanning 
tunneling microscopy as the ideal technique to fabricate such systems and probe their topological properties. 
\end{abstract}
\maketitle

The interest in topological quantum computing and non-abelian braiding has inspired many recent proposals to 
create Majorana fermions (MFs) in various experimental systems. Following  Kitaev's seminal proposal \cite{kitaev_phys2001}, many 
approaches have been considered including those based on topological insulators \cite{PhysRevLett.100.096407, AkhmerovPRL2009}; atoms trapped in 
optical lattices \cite{PhysRevLett.103.020401,DiehlNatPhys2011,PhysRevLett.106.220402}; semiconductors with strong spin-orbit interaction in two and one dimension \cite{SauPRL2010, lutchyn_prl2010, oreg_prl2010}; coupled quantum dots \cite{Jau_natcomm2012, Oreg2012a}; and those that combine magnetism of and superconductivity \cite{ choy_prb2011, GangadharaiahPRL2011,martin_prb2012,PhysRevB.85.020503}. The aim of these approaches is to create a topological superconductor 
in which MFs emerge as the single excitations at the boundaries. Since MFs are its own antiparticles, they are predicted to appear in tunneling 
spectroscopy experiments as zero bias peaks \cite{PhysRevLett.103.237001, PhysRevB.82.180516, Sauprb2010,WimmerNJP2011}. Such peaks have been indeed 
observed in several experiments and interpreted as the signatures of MFs \cite{Mourik25052012, Heiblum_inasmf2012, Xu_LundNanoLett2012}. However, these 
experiments are not spatially resolved to detect the position of the MFs. Additionally, in many instances, the presence of disorder 
can result in spurious zero bias anomalies even when the system is not topological \cite{LiuPRL2012,pikulin_arxiv2012}. It is therefore desirable to identify
easy-to-fabricate condensed matter systems in which MF can be spatially resolved and distinguishable from spurious disorder effects.

In this letter, we theoretically investigate conditions for which a chain of magnetic atoms on the surface of an s-wave superconductor 
can host MF modes. We explore the parameter space for which this system is topological and  show that even relatively short chains 
made of only $\sim 50$ atoms can host robust localized MFs. Our proposed structures can be fabricated using scanning tunneling microscopy (STM), 
which has previously been used to assemble structures of various shapes with tens of atoms using lateral atomic manipulation techniques 
\cite{CrommieScience1993, PhysRevLett.92.056803, Nilius13092002}. Spatially resolved STM spectroscopy of such disorder-free 
chains can be used to probe the presence of MF end modes. 

As shown in Fig. 1, we consider an array of magnetic atoms (such as atoms of 3d or 4f metals with a net magnetic moment) which are deposited 
on a single crystal surface of an s-wave superconductor (such as niobium (Nb) or lead (Pb)) and arranged into chains using the STM. 
The interaction of a single magnetic moment with the superconductor gives rise to the so-called Yu-Shiba-Rusinov states \cite{Yu,Shiba,Rusinov,RevModPhys.78.373}  that have 
been previously detected from both 3d and 4f atoms on the surface of Nb and Pb using an STM \cite{Yazdani21031997, PhysRevLett.100.226801}. 
The results of these previous experiments (with Gd and Mn deposited on Nb) agree well with model calculations in which the magnetic moment is assumed 
to be static \cite{Yazdani21031997, FlattePRL1997, SalkolaPRB}. In addition, recent spin polarized STM studies indicate  that in magnetic arrays with $\gtrsim 10$ atoms spin dynamics is greatly suppressed \cite{Loth13012012}. It is therefore reasonable to model moments of magnetic atoms as static classical spins. 
In general, magnetic moments in these chains can form various configurations including a spiral \cite{PhysRevLett.108.197204}.

\begin{figure}[htb]
\centering
\includegraphics[width=0.5\textwidth]{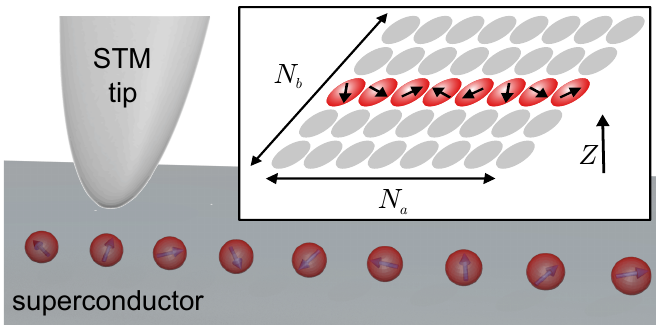}
\caption{Schematic of the experimental setup. An array of magnetic atoms (red spheres) is assembled using scanning tunneling microscope on the surface of 
s-wave superconductor (gray background). The system is modeled by the two dimensional $N_a\times N_b$ array in which 
magnetic atoms are embedded (inset). Throughout the paper we consider the case where magnetic moments are in the plane defined by $N_a$ and $Z$ direction. }
\label{fig:figure1}
\end{figure}

To describe this system we use a two-dimensional tight-binding model Hamiltonian of an s-wave superconductor with an array of magnetic atoms :
\begin{eqnarray}
H =  \sum_{\text{<i,j>} \alpha} (t f_{i \alpha}^\dagger f_{j \alpha} + h.c.)  - \mu \sum_{i\alpha} f_{i\alpha}^\dagger f_{i\alpha} +  \nonumber \\  +
 \sum_{n\alpha \beta} (\vec{B}_n \cdot \vec{\sigma})_{\alpha \beta} f_{n \alpha}^\dagger f_{n \beta} + \sum_i (\Delta_i f_{i\uparrow}^\dagger f_{i\downarrow}^\dagger + h.c).
\end{eqnarray} 
The operators $f$ and $f^\dagger$ correspond to electron annihilation and creation respectively, $t$ is the hopping amplitude between adjacent sites <i,j> of a \emph{two-dimensional} lattice, $\mu$ is the chemical potential, $\Delta_i$ is the local superconducting gap associated with a host superconductor (equal to $\Delta_0$ in the absence of magnetic atoms). The effective magnetic field $\vec{B}_n$ gives rise to a local Zeeman energy on the atoms which are arranged in a \emph{one-dimensional} array of sites \emph{\{n\}}. We consider the case of identical atoms, i.e. $|\vec{B}_n|=B$. Throughout the paper we normalize all simulation parameters to the value of  $\Delta_0$.

In order to obtain the two-dimensional gap profile in the vicinity of the atomic chain, we self-consistently solve the resulting Bogoliubov-de Gennes equations (BdG)  \cite{VieiraPRB2007}. We assume a constant on-site pairing coupling $V$  for a grid of $N_a \times N_b$ lattice sites in the middle of which $N_a$ local magnetic moments with strength $B$ are embedded (see \cite{supp} section 1 for details). The calculations are performed with open boundary conditions (BC) in the $N_b$ direction, and both open and periodic BC in the $N_a$ direction to show the presence or absence of MF at the end of the chain and to compute the Pfaffian index (Pf) \cite{kitaev_phys2001}.  Previous calculations showed that 
a single magnetic moment gives rise to a state inside the superconducting gap that has an energy close to $\Delta_0$ for low $B$. As the value of $B$ is increased the energy of this state is continuously tuned to zero \cite{SalkolaPRB,FlattePRL1997, Morr_prb2003}. This zero crossing is a signature of a quantum phase transition, at which the impurity site traps a single quasi-particle \cite{SodaPTP1967, Shiba}. A similar phase transition occurs in the case of a few magnetic moments \cite{PhysRevB.73.224511,Morr_prb2003}. The transition obviously coincides with a change of the sign of the Pfaffian (computed in a periodic geometry) for the system, indicating a change of the fermion parity in the ground state. This is the characteristic signature of a topological non-trivial phase with MF end modes \cite{kitaev_phys2001}.

An example of a transition into a topologically non-trivial phase for our atomic chain is illustrated in Fig. 2, which shows the lowest energy level and the Pfaffian as a function of $B$ in the case of $96$ magnetic moments. The angle between adjacent magnetic moments, $\theta$, plays a key role in determining whether this system is topological (see below), and has been assumed to be $2\pi/3$ for the results shown in Fig. 2. The most important feature of this calculation is that in the parameter window $2.2< B/\Delta_0 <3.45$, in which for periodic BC in the $N_a$ direction the Pfaffian is negative, and the spatial extent of the lowest excited state (Fig. 2b) (for open BC) shows the presence of MFs at the ends of the chain. This behavior can be contrasted with that of $B/\Delta_0=2.1$ (Fig. 2c). In this case Pfaffian is positive and the lowest energy excitation is distributed approximately evenly along the chain. A  calculation of the local density of states (LDOS) as a function of energy shown in Fig. 2d clearly demonstrates that the topological case shows a zero bias peak associated with MF when tunneling at the end of the chain, while the middle of the system exhibits a mini-gap. In the non-topological phase sufficiently far away from the transition point, the system shows a clear gap throughout the chain and absence of zero energy end modes (Fig. 2e). 

The emerging MF end modes considered here are localized on a very short length scale at the last few sites of the atomic chain. This situation can be contrasted to the proposals involving semiconductor nanowires in proximity with superconductors, where the coherence length of the superconductor sets the length scale for MFs \cite{oreg_prl2010}. The spatial extent of our MFs is reminiscent of the extent of the Yu-Shiba-Rusinov states created by single atoms, which have been shown both experimentally and theoretically to decay on length scales associated with the Fermi wavelength of a superconductor \cite{Yazdani21031997, SalkolaPRB}. Note that these states do have long tails associated with the superconducting coherence length, however this decay is strongly enhanced with an algebraic decay pre-factor \cite{SalkolaPRB,FlattePRL1997}. 

\begin{figure}[htb]
\centering
\includegraphics[width=0.5\textwidth]{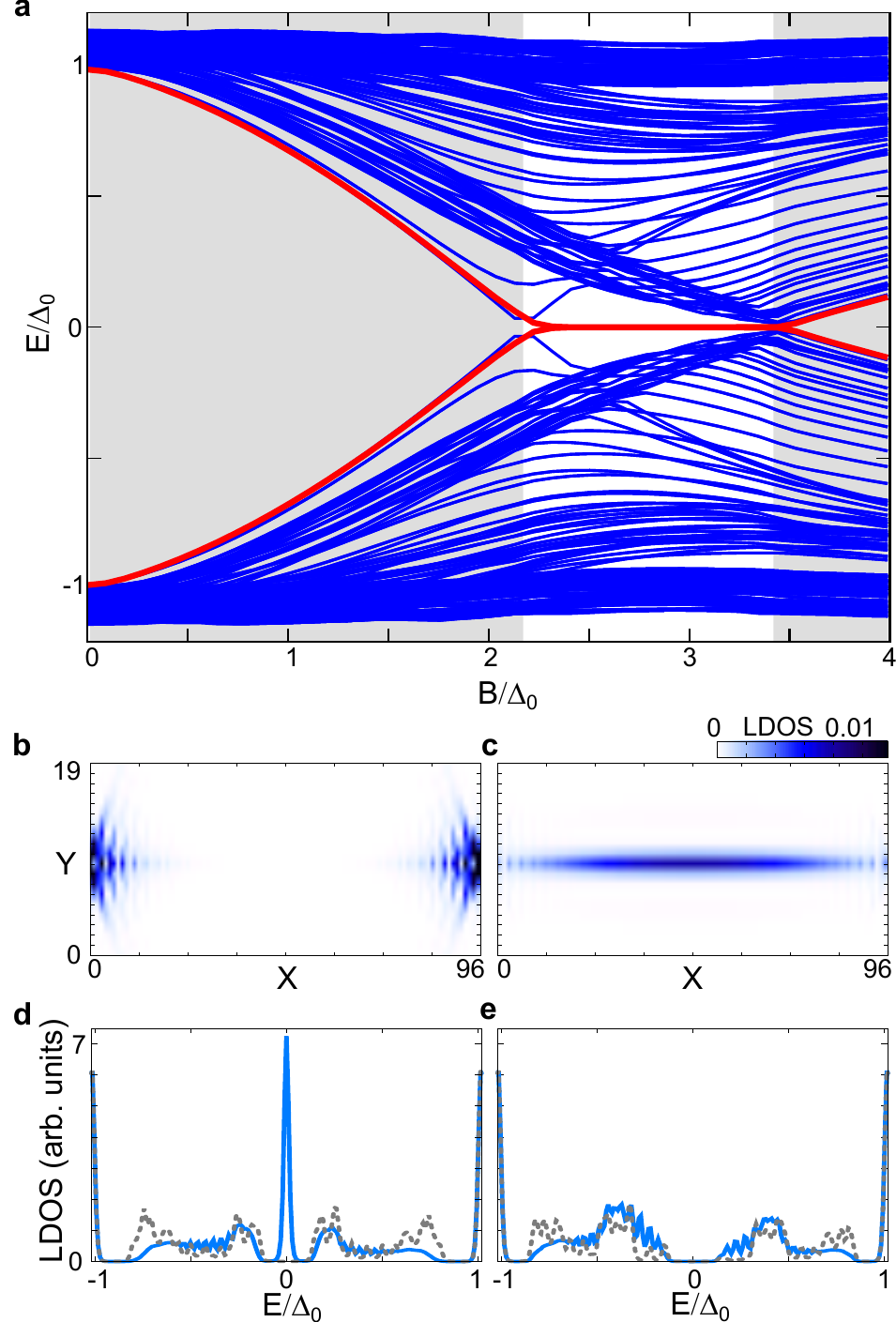}
\caption{
(a)  Calculated energy spectrum, marked by blue lines, for 96 classical spins placed in the middle of the $N_a \times N_b = 96 \times 19$ grid using periodic BC. Parameters for the plot are: $\mu/\Delta_0=2.12$, $t/\Delta_0=2.34$, $V/\Delta_0=2.81$ and $T/\Delta_0=0.01$. The regions 
corresponding to the trivial phase (Pf>0) are shaded gray. Red thick line represents the lowest energy excitation using open BC. (b,c) The spatial distribution of the local density of states corresponding to the lowest excitation state in the non-trivial ($B/\Delta_0$ = 2.87, Pf < 0) and the trivial ($B/\Delta_0 $= 2.23, Pf >0) phase. Lattice coordinates $X$ and $Y$ correspond to the $N_a$ direction (along the chain) and $N_b$ direction (orthogonal to the chain) respectively. (d,e) Local density of states at the chain
ends (blue solid line) and in the middle of the chain (gray dashed line) as a function of energy for non-trivial and trivial phase taking into account first 96 energy eigenvalues. The intrinsic line-width of the energy eigenstates is taken to be $\omega/\Delta_0=1\times10^{-3}$ for this plot. 
}
\label{fig:figure2}
\end{figure}

While we used a self-consistent BdG calculation for realistic modeling of experimental situation, a more efficient approach to gain physical insight into this system is to consider an effective 1D model of magnetic atoms on superconducting sites, which is just the $N_b=1$ limit of our 2D model. Note that in 1D, all information about the superconductor is simply included in the strength of the on-site $s$-wave gap $\Delta_0$ and the hopping term describes coupling between the impurities on superconducting sites only (as opposed to the superconductor bandwidth in BdG model above, see \cite{supp} section 2).  Fig. 3 shows that a 1D model qualitatively gives similar results the 2D model. Importantly,  the hopping term, which can be tuned experimentally by placing atoms at different distances, may also drive quantum phase transition from the trivial phase (Pf>0) to the topological phase (Pf<0) with MFs at the ends.  A one-dimensional version of this Hamiltonian is also considered in Ref. \cite{choy_prb2011} in the context of MFs in disordered magnetic islands on a superconductor.
\begin{figure}[htb]
\centering
\includegraphics[width=0.5\textwidth]{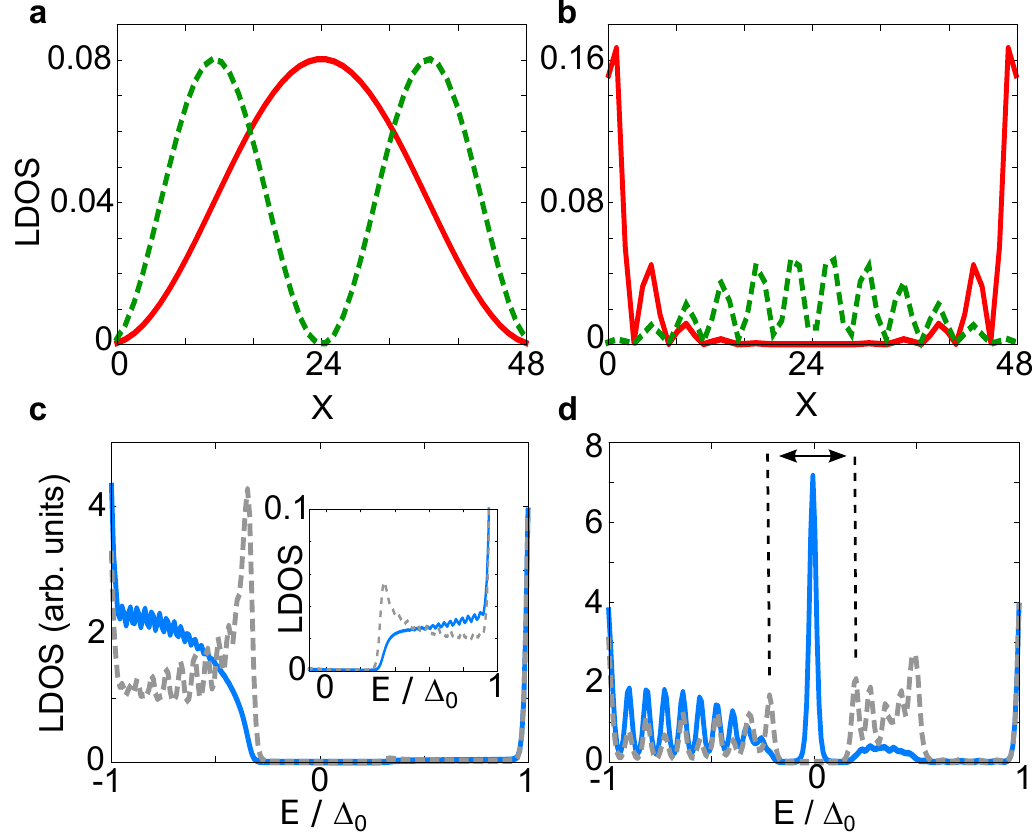}
\caption{(a,b) The spatial profile of the two lowest excitation states  of magnetic chain containing 48 atoms for $\mu/\Delta_0=4$, $B/\Delta_0=5$, $\theta=\pi/2$ (marked by red solid and green dashed line respectively). Tuning the hopping term $t$ drives quantum phase
 transition from the trivial ($t/\Delta_0=0.4$, Pf > 0) (a) to the topological ($t/\Delta_0=1$, Pf <0) phase (b).  (c,d), Local density of states calculated for the same parameters as in (a) and (b) at the chain ends (blue solid line) and in the middle of the chain (gray dashed line). Note that for this choice of parameters spectrum in (c) is asymmetric in energy (see inset). Importantly, in (d) the two MF states  around zero energy are separated by the effective mini gap $\Delta_p$ from the other states in the spectrum (marked by double arrow line).}
\label{fig:figure3}
\end{figure}

A key advantage of the 1D model is that it lends itself to an analytical solution, which shows that for a given angle $\theta$ between adjacent moments, the 
Pfaffian for the system is negative when                                                                                                                                                                                                                                                                                                                                                                                                                                                                                                                                                                                                                                                                                                                                                                                                                                                                                                                                                                                                                                                                                                                                                                                                                                                                                                                                                                                                                                                                                                                                                                                                                                                                                                                                                                                                                                                                                                                                                                                                                                                                                                                                                                                                                                                                                                                                                                                                                                                                                                                                                                                                                                                                                           
\begin{eqnarray}
 \sqrt{\Delta_0^2+  (|\mu| + 2 |t\cos(\theta/2)|)^2}> |B|, \nonumber \\
|B| >\sqrt{\Delta_0^2+  (|\mu| - 2 |t\cos(\theta/2)|)^2} \label{condition1}
\end{eqnarray}
(see \cite{supp}, section 3 for the derivation).   
The negative value of the Pfaffian is a necessary condition for this system to be in a topological phase; however, it not sufficient, as  the bulk of atomic chain remains must also be gapped.  For example, $\theta=0, \pi$ have the widest range of negative Pfaffian in Eq. \ref{condition1}; unfortunately, this full range is gap-less. The min-gap for low energy excitation is related to strength of the $p$-wave pairing that emerges on the chain because of the combination of  hopping, pairing, and local Zeeman terms in the Hamiltonian. Calculations of the spectra in both 2D and 1D model described above reveal the energy scale, which separates the zero energy MF states (localized at the two ends) from the next available excitation of the system. In a certain limit, the 1D model can be directly mapped \cite{choy_prb2011} to the original proposal by Kitaev for realization of MF end mode, which is a superconducting wire with nearest neighbor pairing \cite{kitaev_phys2001}, but general eigenvalues can be obtained even without this mapping, see \cite{supp} section 2. The value of this mini-gap depends on the relative values of $\mu$, $t$, $B$, and angle $\theta$ (see Fig 4). 

\begin{figure}[htb]
\centering
\includegraphics[width=0.5\textwidth]{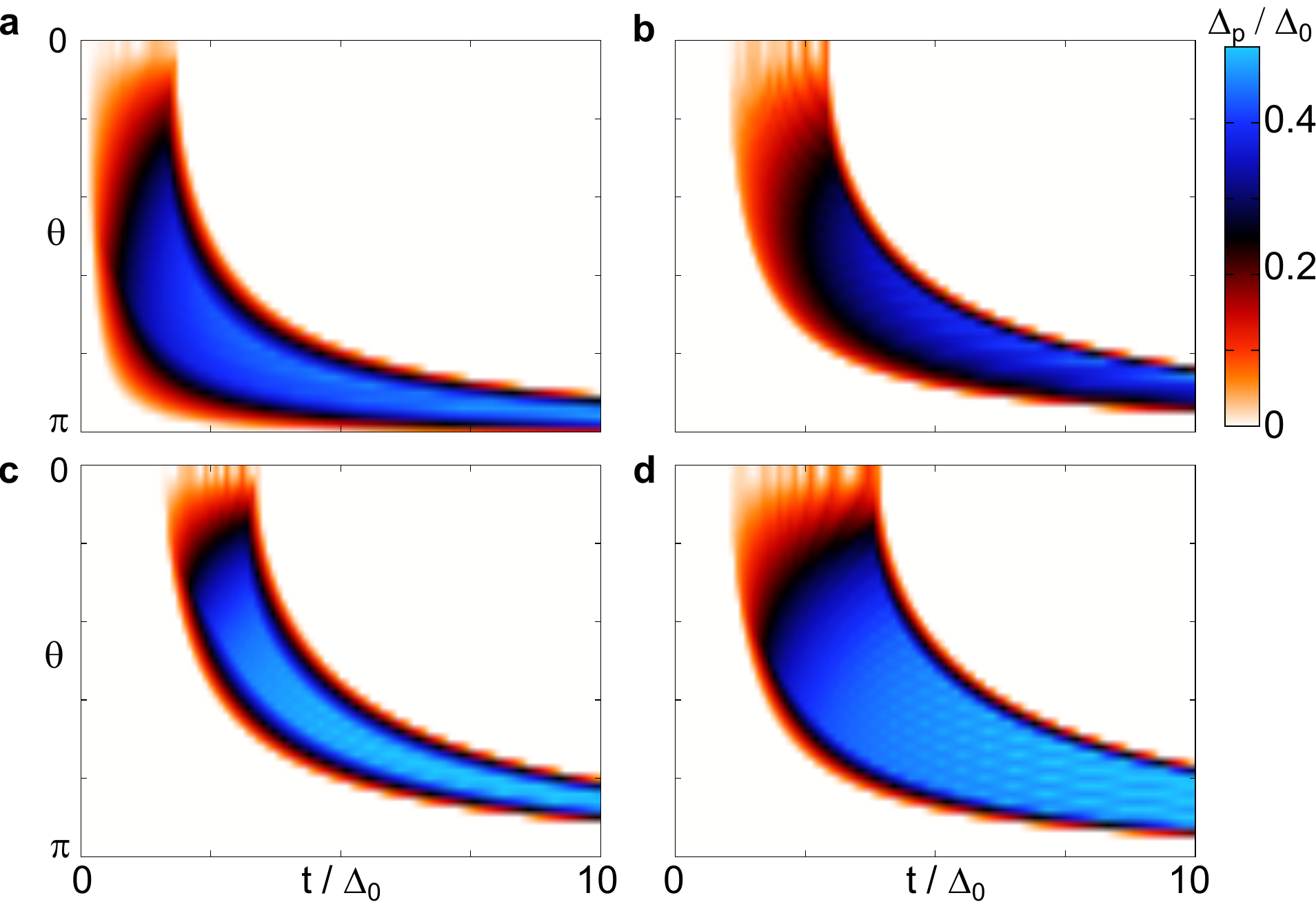}
\caption{The value of the mini gap as a function of tunnel coupling and angle $\theta$ calculated for the 1D model : (a)  $\mu/\Delta_0=2$ and $B/\Delta_0=3$; (b) $\mu/\Delta_0=2$ and $B/\Delta_0=5$ ;  (c) $\mu/\Delta_0=5$ and $B/\Delta_0=2$; (d) $\mu/\Delta_0=5$ and $B/\Delta_0=4$. }
\label{fig:figure4}
\end{figure}

A non-collinear arrangement of magnetic moments in a chain is essential to realize robust MF end modes. When transformed to a basis parallel to the spiraling on-site magnetic field, the hopping becomes spin-dependent giving rise to spin-orbit coupling and hence to the usual mechanisms for MF end modes.  Without detailed modeling of the surface magnetism it is difficult to predict whether specific magnetic atomic chains would have a spiral spin-arrangement. We suggest that exploring the full freedom of the  linear chain geometry may provide a feasible approach to create favorable conditions for non-collinear magnetic 
moments of adjacent atoms. For example, double or zig-zag chain structures with anti-ferromagnetic interactions are likely to become frustrated and result in spiral orientation of magnetic moments in the chain \cite{PhysRevLett.108.197204}. To explore some of these possible geometries (Fig. 5a), we map these chains into equivalent 
linear chains with the nearest $t_1$ and the next nearest $t_2$ hopping  as shown 
in Fig. 5b. In the simplest case for which $\theta$ is assumed constant, we show that these chains can also support topological phase when 
\begin{eqnarray}
\sqrt{\Delta_0^2+  (\mu + 2 \cos(\theta/2) t_1 - 2 \cos(\theta) t_2)^2}> |B|, \nonumber \\
|B|>\sqrt{\Delta_0^2+  (\mu - 2 \cos(\theta/2) t_1 - 2 \cos(\theta) t_2)^2} \label{condition2}
\end{eqnarray}
 (see \cite{supp} section 3a for further details).  We note again that these chains may provide easy-to-fabricate structures that would ensure 
non-collinear spin arrangements required for realization of MF end modes.

\begin{figure}[htb]
\centering
\includegraphics[width=0.5\textwidth]{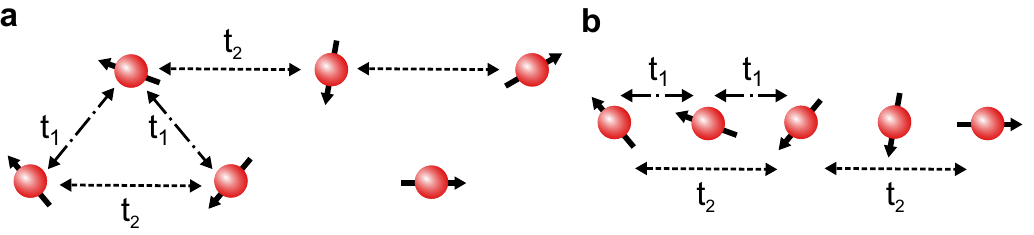}
\caption{(a) Array of magnetic atoms arranged in two rows (zig-zag chain). The coupling among neighboring atoms 
corresponding to different rows is $t_1$ and the coupling between atoms within the same row is $t_2$. (b)
Equivalent magnetic moment configuration represented as a single chain with the next nearest coupling.}
\label{fig:figure5}
\end{figure}

Lastly, we comment on the experimental feasibility of the proposed approach. As shown here the strength of 
the mini-gap associated with the $p$-wave pairing can sometimes exceed 30-40$\%$ of the gap of the host superconductor (Fig. 4). Nevertheless, using 
an s-wave superconductor with large gap $\Delta_0$ (and measuring at the lowest temperatures) would increase the chance of experimental 
success. Other factors such as size of the magnetic moment $B$ or hopping matrix element $t$ are also important
and can be optimized experimentally using magnetic atoms with different spin or building chains with different spacing. 
A systematic experimental approach can start by characterizing the single-impurity states and their modification when
impurities are brought close enough to interact \cite{PhysRevLett.100.226801}. These measurements could be used to map 
effective 1D model parameters (effective hopping, chemical potential and exchange coupling) and allow investigation of the finite size effects 
on the excitation spectrum. A different approach would be to start from magnetic 
chains grown using self-assembled techniques. Note that self-assembled chains consisting of $\sim 50$ atoms with spiral arrangement of magnetic 
moments are already reported \cite{PhysRevLett.108.197204}. Such chains would be an ideal starting point to investigate interaction 
between Majorana fermions. For example, examining coupled chains can provide direct experimental means to demonstrate the $Z_2$ character 
of the MF end modes by showing that they appear only in odd number of coupled chains. Finally, as structures of different shapes are equally easy to assemble 
in STM,  one can envision viable route towards braiding experiments in arrays of coupled chains in a similar fashion as proposed for 
semiconductor nanowire structures \cite{alicea_natphys2011, PhysRevB.85.144501, beenakker2012a}. 

We thank R. Lutchyn and J. Seo for the discussions. This work is supported by NSF-DMR1104612 and NSF-MRSEC programs through the Princeton Center for Complex Materials (DMR-0819860), ONR, ARO, and DARPA-SPAWAR grant N6601-11-1-4110. S. N-P acknowledges support of of the European Community under a Marie-Curie OEF fellowship. B. A. B. acknowledges support of the programs  CAREER DMR-095242, ONR-N00014-11-1-0635, ARMY-245-6778 and the Packard Foundation.

\bibliography{mf_nourl}
\onecolumngrid
\clearpage
\appendix


\section*{Supplementary material}

# Supplementary information "Majorana fermions in chains of magnetic atoms on a superconductor"

S. Nadj-Perge, I. K. Drozdov, B. A. Bernevig and Ali Yazdani

*Department of Physics, Princeton University, Princeton, New Jersey 08544*

(Dated: March 25, 2013)

## 1. DETAILS OF THE TWO-DIMENSIONAL MODEL

The numerical calculations of the two dimensional model are preformed self-consistently [S1]. First we solve Bogoliubov de Gennes equation for the Hamiltonian of our system (Eq. 1) and find electron- and hole- like wave functions $u_{n,\sigma}(i,j)$ and $v_{n,\sigma}(i,j)$. The excitation spectrum is then used to determine local on-site value of superconducting gap $\Delta(i,j)$

$$\Delta(i,j) = -V/2 \sum_n \{(2f_n - 1)(u_{n,\uparrow}(i,j)^* v_{n,\downarrow}(i,j) + u_{n,\downarrow}(i,j)v_{n,\uparrow}^*(i,j))\} \tag{S1}$$

where $V$ is the pairing strength, $f_n = 1/(e^{\epsilon_n/T} + 1)$ Fermi function, $n$ sums over energy eigenvalues $\epsilon_n$ and $T$ is the temperature. In the next iteration the obtained values $\Delta(i,j)$ are used to recalculate the excitations spectrum. This iterative process is stopped after 20 iterations, typically assuring relative error of $\Delta(i,j)$ to be less than $10^{-2}$. In the last step we also calculate the spatial profile of local density of states (LDOS) defined as

$$\rho(i,j,\epsilon) = \sum_{n,\sigma} (|u_{n,\sigma}(i,j)|^2 \delta(\epsilon - \epsilon_n) + |v_{n,\sigma}(i,j)|^2 \delta(\epsilon + \epsilon_n)).$$

We used open boundary conditions for the $N_b$ direction and periodic and open boundary conditions in the $N_a$ direction. Periodic boundary conditions are used to compute Pfaffian index of the system (as discussed in the section 2). When the Pfaffian is negative (i.e. in topological phase) open boundary conditions are used to verify the existence of the two energy zero modes with LDOS localized at the chain ends. These two states are identified as a pair of Majorana fermions (MFs).

Although self-consistent calculations are useful to explicitly verify the occurrence of MFs in more realistic conditions, they require intensive computational resources. Qualitatively similar results can be obtained in a toy model without imposing self-consistency. This type of calculations allow to quickly investigate the parameter space for which the system is in topological phase. Further, in the one-dimensional case (for $N_b = 1$) analytical conditions can be found in some cases as shown in the following sections.

## 2. HAMILTONIAN IN ONE DIMENSION AND THE TOPOLOGICAL INDEX

In the case of $N_b = 1$, the Hamiltonian in Eq. 1 reduces to:

$$H = \sum_{n\alpha} t_n f_{n\alpha}^\dagger f_{n+1\alpha} + t_n^* f_{n+1\alpha}^\dagger f_{n\alpha} - \mu \sum_{n\alpha} f_{n\alpha}^\dagger f_{n\alpha} + \sum_{n\alpha\beta} (\vec{B}_n \cdot \vec{\sigma})_{\alpha\beta} f_{n\alpha}^\dagger f_{n\beta} + \sum_n \Delta_0 f_{n\uparrow}^\dagger f_{n\downarrow}^\dagger + \Delta_0 f_{n\downarrow} f_{n\uparrow} \tag{S2}$$

In this one-dimensional Hamiltonian we assume a fixed value of $\Delta_0$ for every site. As noted in the main text, for this model, tunnel coupling $t$ defines solely hopping term between magnetic atoms. Note that this is an experimentally tunable parameter since it can be changed by placing atoms on different distances. To define Pfaffian of this system, we first rewrite the Hamiltonian using the real-fermion Majorana representation. The mapping from complex fermions to MF basis can be done in the following way:

$$a_{2n-1,\sigma} = f_{n\sigma} + f_{n\sigma}^\dagger, \quad a_{2n,\sigma} = -i(f_{n\sigma} - f_{n\sigma}^\dagger). \qquad a_{n\sigma}^\dagger = a_{n\sigma}, \quad \{a_{n\sigma}, a_{m\sigma'}\} = 2\delta_{nm}\delta_{\sigma,\sigma'}. \tag{S3}$$

Note that the site index of the complex fermion $f$ goes from 1 to $N_a$ while the index of the real Majorana fermions goes from 1 to $2N_a$. Since for each complex fermion $f_{n\sigma}$ there is two real MFs $a_{2n\sigma}$ and $a_{2n-1\sigma}$ the number of sites is effectively doubled. In Majorana representation 1D Hamiltonian reads as

$$H = \sum_{n=1}^N \frac{iRe(t_n)}{2}(a_{2n-1\sigma}a_{2n+2\sigma} - a_{2n\sigma}a_{2n+1\sigma}) + \frac{iIm(t_n)}{2}(a_{2n-1\sigma}a_{2n+1\sigma} + a_{2n\sigma}a_{2n+2\sigma}) -$$
$$- \frac{i\mu}{2}a_{2n-1\sigma}a_{2n\sigma} + \frac{iB_{nz}}{2}(a_{2n-1\uparrow}a_{2n\uparrow} - a_{2n-1\downarrow}a_{2n\downarrow}) + \frac{iB_{nx}}{2}(a_{2n-1\uparrow}a_{2n\downarrow} - a_{2n\uparrow}a_{2n-1\downarrow})$$
$$- \frac{iB_{ny}}{2}(a_{2n-1\uparrow}a_{2n-1\downarrow} + a_{2n\uparrow}a_{2n\downarrow}) + \frac{i\Delta_0}{2}(a_{2n-1\downarrow}a_{2n\uparrow} + a_{2n\downarrow}a_{2n-1\uparrow}) \tag{S4}$$

We have assumed a real order parameter $\Delta_0 = \Delta_0^*$. Importantly, Hamiltonian in this representation can be rewritten as

$$H = \frac{i}{4} \sum_{l,m=1}^{2N} a_{l\sigma} A_{l\sigma;m\sigma'} a_{m\sigma'} \tag{S5}$$

where $A_{l\sigma;m\sigma'}^* = A_{l\sigma;m\sigma'} = -A_{m\sigma';l\sigma}$ is an $4N_a \times 4N_a$ ($2N_a$ MFs per spin species) antisymmetric matrix with elements easily readable from the above expression. For the upper defined form, we define the topological index as

$$\text{sign}(Pf(\frac{i}{4} A_{l\sigma;m\sigma'})) \tag{S6}$$

which effectively describes the topological phase of the system [S2]. Any change between $+1$ and $-1$ of the index indicates a phase transition where the system changes its topological phase. Note that the extension of this index to the 2D model is straightforward and was done using the same majorana representation. Starting from a $(2 \times N_a \times N_b) \times (2 \times N_a \times N_b)$ matrix representation of the Hamiltonian (Eq. 1) in the complex fermion basis we get $(4 \times N_a \times N_b) \times (4 \times N_a \times N_b)$ Hamiltonan matrix in Majorana basis. The corresponding $A_{l\sigma;m\sigma'}$ matrix is again antisymmetric and the same definition of the topological index can be used.

### 3. ANALYTICAL SOLUTION FOR TOPOLOGICAL PHASE IN THE CASE OF SPIN HELIX

In this section we solve the one-dimensional Hamiltonian (Eq. S2) for periodic boundary conditions when angle $\theta$ between adjacent magnetic atoms is constant. Since all atoms are identical the magnetic moments on each site have same magnitude but differ in direction: $\vec{B}_n = B_0(\sin(\theta_n)\cos(\phi_n)\hat{x} + \sin(\theta_n)\sin(\phi_n)\hat{y} + \cos(\theta_n)\hat{z})$. We start from the complex fermion representation and first rotate the fermionic operators into an $\uparrow\downarrow$ basis on each atomic site using unitary rotation described in Ref. [S3]

$$\begin{pmatrix} f_{n\uparrow} \\ f_{n\downarrow} \end{pmatrix} = U_n \begin{pmatrix} g_{n\uparrow} \\ g_{n\downarrow} \end{pmatrix} = \begin{pmatrix} \cos(\theta_n/2) & -\sin(\theta_n/2)e^{-i\phi_n} \\ \sin(\theta_n/2)e^{i\phi_n} & \cos(\theta_n/2) \end{pmatrix} \begin{pmatrix} g_{n\uparrow} \\ g_{n\downarrow} \end{pmatrix} \tag{S7}$$

As $U_n$ is a unitary transformation fermions $g_{n\sigma}$ have the same anticommutation properties as $f_{n\sigma}$. This transformation doesn't change the form of the chemical potential and gap terms, whereas the magnetic moment term becomes diagonal. The hopping term now acquires inter-spin components, and is no longer diagonal in spin. The rotated Hamiltonian now reads as

$$H = \sum_{n,\alpha,\beta} t_n \Omega_{n,\alpha,\beta} g_{n\alpha}^\dagger g_{n+1\beta} + t_n^* \Omega_{n,\beta,\alpha}^* g_{n+1\alpha}^\dagger g_{n\beta} + B_0 \sigma_{z\alpha\beta} g_{n\alpha}^\dagger g_{n\beta} - \mu \sum_{n\alpha} g_{n\alpha}^\dagger g_{n\alpha} + \sum_n \Delta_0(g_{n\uparrow}^\dagger g_{n\downarrow}^\dagger + g_{n\downarrow} g_{n\uparrow}) \tag{S8}$$

where $\Omega_n$ is defined as

$$\Omega_n = U_n^\dagger U_{n+1} = \begin{pmatrix} \alpha_n & -\beta_n^* \\ \beta_n & \alpha_n^* \end{pmatrix} \tag{S9}$$

with matrix elements being $\alpha_n = \cos(\theta_n/2)\cos(\theta_{n+1}/2) + \sin(\theta_n/2)\sin(\theta_{n+1}/2)e^{-i(\phi_n - \phi_{n+1})}$ and $\beta_n = -\sin(\theta_n/2)\cos(\theta_{n+1}/2)e^{i\phi_n} + \cos(\theta_n/2)\sin(\theta_{n+1}/2)e^{i\phi_{n+1}}$. In the case of magnetic moments aligned in the x-z plane with constant angle $\theta$ between adjacent moments coefficients $\alpha_n$ and $\beta_n$ reduce to $\alpha = \cos(\theta/2)$ and $\beta = \sin(\theta/2)$ respectively. We rewrite the rotated operators $g_{n\sigma}$ in Majorana basis as follows:

$$g_{n\sigma} = \frac{1}{2}(b_{2n-1\sigma} + ib_{2n\sigma}), \quad g_{n\sigma}^\dagger = \frac{1}{2}(b_{2n-1\sigma} - ib_{2n\sigma}). \tag{S10}$$

The $b$ Majorana operators are the rotated $a$ Majorana operators in the basis parallel to the magnetic field. In this basis, the Hamiltonian becomes:

$$
\begin{aligned}
H = &\sum_n \frac{t_n(\alpha_n - \alpha_n^*)}{4}(b_{2n-1\uparrow}b_{2n+1\uparrow} + b_{2n\uparrow}b_{2n+2\uparrow} - b_{2n-1\downarrow}b_{2n+1\downarrow} - b_{2n\downarrow}b_{2n+2\downarrow}) + \\
&+ \frac{it_n(\alpha_n + \alpha_n^*)}{4}(b_{2n-1\uparrow}b_{2n+2\uparrow} - b_{2n\uparrow}b_{2n+1\uparrow} + b_{2n-1\downarrow}b_{2n+2\downarrow} - b_{2n\downarrow}b_{2n+1\downarrow}) + \\
&+ \sum_n \frac{t_n(\beta_n - \beta_n^*)}{4}(b_{2n-1\uparrow}b_{2n+1\downarrow} + b_{2n\uparrow}b_{2n+2\downarrow} - b_{2n-1\downarrow}b_{2n+1\uparrow} - b_{2n\downarrow}b_{2n+2\uparrow}) + \\
&+ \frac{-it_n(\beta_n + \beta_n^*)}{4}(b_{2n-1\uparrow}b_{2n+2\downarrow} - b_{2n\uparrow}b_{2n+1\downarrow} + b_{2n-1\downarrow}b_{2n+2\uparrow} + b_{2n\downarrow}b_{2n+1\uparrow}) + \\
&+ \sum_n \frac{iB_n}{2}(b_{2n-1\uparrow}b_{2n\uparrow} - b_{2n-1\downarrow}b_{2n\downarrow}) - \frac{i\mu}{2}(b_{2n-1\uparrow}b_{2n\uparrow} + b_{2n-1\downarrow}b_{2n\downarrow}) + \frac{i\Delta_0}{2}(b_{2n-1\downarrow}b_{2n\uparrow} - b_{2n-1\uparrow}b_{2n\downarrow})
\end{aligned} \tag{S11}
$$

To explicitly calculate the Pfaffian index and energy eigenvalues we need to Fourier transform MF operators as

$$b_{2n-1\sigma} = \frac{1}{\sqrt{N}}\sum_n e^{-iqn} b_{q,1,\sigma}, \quad b_{2n\sigma} = \frac{1}{\sqrt{N}}\sum_n e^{-iqn} b_{q,2,\sigma} \tag{S12}$$

where $1, 2$ are the two flavors of MF operators coming from splitting the on-site fermion, and $q$ is the Fourier vector which takes values to be determined below (there are several cases for the values of $q$ depending on the boundary conditions, which in turn depend on the value of the rotation of the spin). We can then write the Hamiltonian in the basis $b_q = (b_{q,1,\uparrow}, b_{q,1,\downarrow}, b_{q,2,\uparrow}, b_{q,2,\downarrow})$:

$$H = \frac{i}{4}\sum_q b_q^\dagger A(q) b_q \tag{S13}$$

where $A(q)$ matrix has nonzero matrix elements:

$$A_{13}(q) = 2t\alpha\cos(q) + B_0 - \mu, \quad A_{14}(q) = 2it\beta\sin(q) - \Delta_0 \tag{S14}$$

$$A_{23}(q) = -2it\beta\sin(q) + \Delta_0 \quad A_{24}(q) = 2t\alpha\cos(q) - B_0 - \mu \tag{S15}$$

Periodic boundary conditions restrain the angle $\theta$ to be $\theta = \frac{2\pi}{N_a}N_h$ where $N_h$ defines number of times the magnetic moment is rotating in an $N_a$ site system. In other words, the magnetic moment rotates by an angle $2\pi N_h$ going from site $n = 1$ to site $n = N_a$. A further complication may arise since we are rotating a spin $1/2$ system. If the magnetic moment rotates an odd number of times, in the rotated basis we have a situation where spin $1/2$ is rotating an odd number of times giving rise to a $-1$ sign. This changes the periodic boundary conditions to anti-periodic. Therefore, the allowed values of the Fourier vector $q$ depend on whether $N_h$ and $N_a$ are odd or even. The four cases with allowed values of $q$ are summarized as follows:

$$q = \begin{cases} \frac{2\pi}{N_a}j & \text{If } \mod[N_h, N_a] \leq N_a/2 \text{ and if } N_h \text{ even} \\ \frac{\pi}{N_a}(2j+1) & \text{If } \mod[N_h, N_a] \leq N_a/2 \text{ and if } N_h \text{ odd} \\ \frac{2\pi}{N_a}j & \text{If } \mod[N_h, N_a] > N_a/2 \text{ and if } \mod[N_a - \mod[N_h, N_a], 2] = 0 \\ \frac{\pi}{N_a}(2j+1) & \text{If } \mod[N_h, N_a] > N_a/2 \text{ and if } \mod[N_a - \mod[N_h, N_a], 2] = 1 \end{cases} \tag{S16}$$

Replacing the allowed $q$ in the expression for Pfaffian one can obtain topological index:

$$\text{Sign}(Pf[A(q=0)])\text{Sign}(Pf[A(q=\pi)]) = \text{Sign}([B_0^2 - (\Delta_0^2 + (\mu + 2\alpha t)^2)][B_0^2 - (\Delta_0^2 + (\mu - 2\alpha t)^2)]). \tag{S17}$$

Note that $q = 0, \pi$ values are allowed only in only when $N_h$ and $N_s$ are both even. The nontrivial topological phase appears when the above index is $-1$ i.e. when:

$$\sqrt{\Delta_0^2 + (|\mu| + 2|\alpha t|)^2} > |B_0| > \sqrt{\Delta_0^2 + (|\mu| - 2|\alpha t|)^2}. \tag{S18}$$

This is the necessary condition for Majorana modes to appear in the one dimensional model assuming constant $\theta$. For actually having Majorana modes the system also has to be gapped. Therefore it is necessary to calculate the energy spectrum which can be easily obtained by diagonalizing the Eq. S13. For each $q$ four eigenvalues are

$$E(q) = \pm\frac{1}{4}\sqrt{B^2 + 4(1-\alpha^2)t^2\sin^2(q) + (\mu - 2\alpha t\cos(q))^2 + \Delta_0^2 \pm 2\sqrt{B_0^2\Delta_0^2 + (\mu - 2\alpha t\cos(q))^2(B_0^2 + 4(1-\alpha^2)t^2\sin^2(q))}} \tag{S19}$$

where all possible combinations of the $\pm$ signs are possible. At $q = 0$, the bands closest to the 0 energy become $\pm(B_0 - \sqrt{\Delta_0^2 + (\mu - 2\alpha t)^2})$ and at $q = \pi$ they are $\pm(B - \sqrt{\Delta_0^2 + (\mu + 2\alpha t)^2})$. This shows that when the Pfaffian changes sign at the edge of the topological phase the gap also vanishes.

Notice that for anti-ferromagnetic coupling ($\theta = \pi$) the criteria in Eq. S18 collapses to a single point i.e for this magnetic moment orientation system never crosses into topological phase. Naively, one could expect that topological phase is maximized for ferromagnetic coupling ($\theta = 0$). In this case the energy eigenvalues (Eq. S19) reduce to $E(q) = \pm B \pm \sqrt{\Delta_0^2 + (\mu - 2t\cos(q))^2}$. Unfortunately, for this case, in the whole range where Pfaffian is negative the system is also gapless and therefore it does not have topological phase.

The absence of MF modes in the ferromagnetic and anti-ferromagnetic configurations can be also confirmed using a self consistent two dimensional model as illustrated by example shown in Fig. S1. In the anti-ferromagnetic

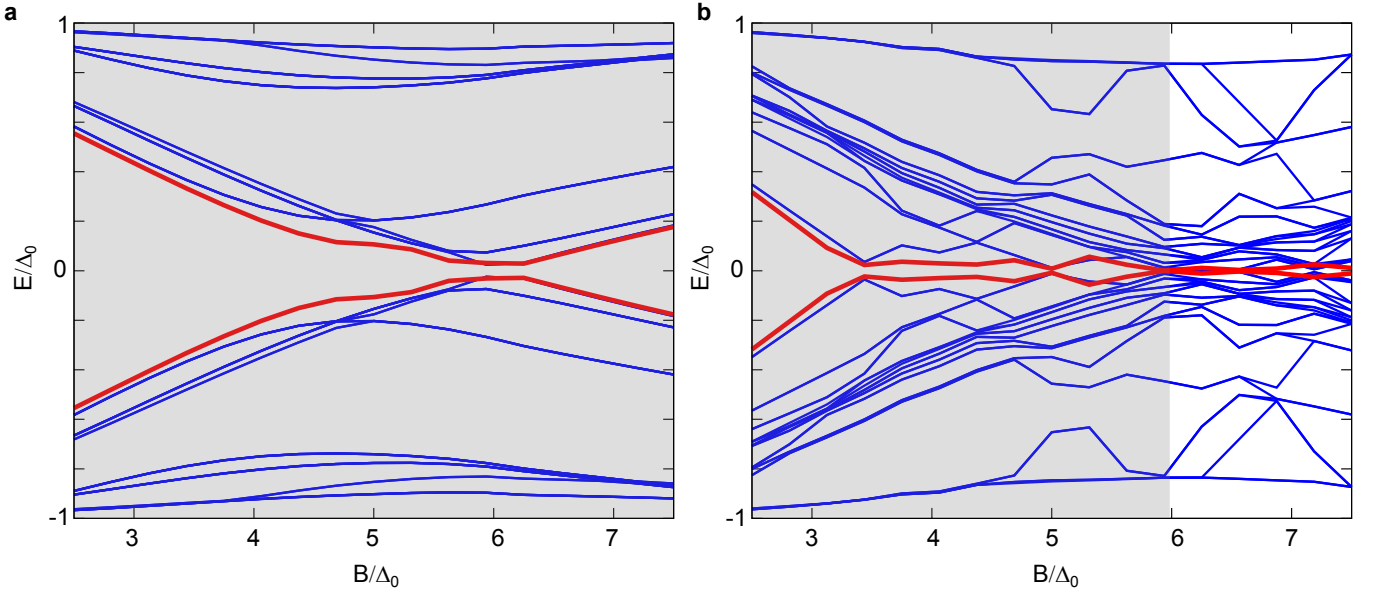

FIG. S1. Energy spectrum showing lowest 24 states of the magnetic chain for anti-ferromagnetic (a) and ferromagnetic (b) orientation of magnetic moments ($N_a \times N_b = 24 \times 25$, $t/\Delta_0 = 3.83$, $\mu/\Delta_0 = 3.19$, $T/\Delta_0 = 0.012$). For anti-ferromagnetic coupling (a) the topological index stays positive ($Pf > 0$) in the whole range of $B$. For ferromagnetic coupling Pfaffian changes sign around $B/\Delta_0 \approx 6$ but the mini-gap in this range (for $B/\Delta_0 > 6$) practically does not exist. Note that at around $B/\Delta_0 \approx 5$ the lowest energy levels also come very close to zero energy, but the Pfaffian does not change sign around this point.

configuration the levels approach zero energy but the two lowest levels never cross, (Fig. S1(a)). Therefore, the chain remains in the trivial phase ($Pf > 0$) for the whole range of B. On the other hand, for the ferromagnetic configuration several levels practically collapse as the lowest energy level crosses zero, Fig. S1(b). In this configuration $Pf < 0$ but there is no well defined gap of the system. In other word system is gapless and the MF modes can not be therefore separated from the remaining states in the spectrum.

### 3a. Topological phase of the simple zig-zag chain

The analytical results above can be extended to more complex chains. In this subsection we consider zig-zag chains of impurities where atoms are placed in two rows as shown in Fig. 5 of the main text. This configuration is interesting since, for antiferromagnetic coupling, it represents a frustrated spin system with potentially ground state where spins are canted off from purely anti-ferromagnetic alignment. A zig-zag chain can be thought of as a skewed ladder, where one of the chains in the ladder is shifted mid-bond with respect to the other chain. In this case, the zig-zag chain can be thought of as a single chain with both nearest neighbour hopping $t_{1n}$ and second neighbour hopping $t_{2n}$. To find the Hamiltonian for we can again project to the on-site spin basis $\vec{B}_n = B_n(\sin(\theta_n)\cos(\phi_n)\hat{x} + \sin(\theta_n)\sin(\phi_n)\hat{y} + \cos(\theta_n)\hat{z})$ with $B_n > 0$ and the Hamiltonian becomes:

$$H = \sum_{n,\alpha,\beta} t_{1n}\Omega_{1n,\alpha,\beta}g_{n\alpha}^\dagger g_{n+1\beta} + t_{2n}\Omega_{2n,\alpha,\beta}g_{n\alpha}^\dagger g_{n+2\beta} + B_n\sigma_{z\alpha\beta}g_{n\alpha}^\dagger g_{n\beta} - \mu\sum_{n\alpha} g_{n\alpha}^\dagger g_{n\alpha} + \sum_n \Delta_0(g_{n\uparrow}^\dagger g_{n\downarrow}^\dagger + g_{n\downarrow}g_{n\uparrow}) \quad \text{(S20)}$$

where the matrix $\Omega_{1n}$ is identical to the one before while:

$$\Omega_{2n} = U_n^\dagger U_{n+2} = \begin{pmatrix} \alpha_{2n} & -\beta_{2n}^* \\ \beta_{2n} & \alpha_{2n}^* \end{pmatrix} \quad \text{(S21)}$$

where $\alpha_{2n} = \cos(\theta_n/2)\cos(\theta_{n+2}/2) + \sin(\theta_n/2)\sin(\theta_{n+2}/2)e^{-i(\theta_n-\theta_{n+2})}$ and $\beta_{2n} = -\sin(\theta_n/2)\cos(\theta_{n+2}/2)e^{i\phi_n} + \cos(\theta_n/2)\sin(\theta_{n+2}/2)e^{i\theta_{n+2}}$.

If we now assume that the magnetic moment form spiral configuration, just like in the single-chain case, the situation is easy to analyze. We take $t_{1n} = t_1$, $t_{2n} = t_2$ constant across the chain. By taking $\phi_n = 0$ and $\theta_n = (n-1)\frac{2\pi}{N_s}N_h$ in

which the spin is rotating $N_h$ times in an $N_s$ site zig-zag chain, we have:

$$\alpha_{1n} = \alpha_1 = \cos(\pi N_h/N_s), \quad \beta_{1n} = \beta_1 = \sin(\pi N_h/N_s), \quad \alpha_{2n} = \alpha_2 = \cos(2\pi N_h/N_s), \quad \beta_{2n} = \beta_2 = \sin(2\pi N_h/N_s)$$
(S22)

By going to the Majorana basis and Fourier space the Hamiltonian can again in the same form as Eq. S13 $H = \frac{i}{4}\sum_q b_q^\dagger A(q) b_q$ with nonzero elements of the matrix $A(q)$ being

$$A_{13}(q) = 2t_1\alpha_1\cos(q) + 2t_2\alpha_2\cos(2q) + B_0 - \mu, \quad A_{14}(q) = 2it_1\beta_1\sin(q) + 2it_2\beta_2\sin(2q) - \Delta_0, \quad (S23)$$

$$A_{23}(q) = -2it_1\beta_1\sin(q) - 2it_2\beta_2\sin(2q) + \Delta_0, \quad A_{24}(q) = 2t_1\alpha_1\cos(q) + 2t_2\alpha_2\cos(2q) - B_0 - \mu. \quad (S24)$$

This result is simple extension the introduction of the second neighbor gives just another added hopping contribution in the Majorana basis. The energy spectrum of the problem reads:

$$E(q) = \pm\frac{1}{4}\sqrt{B_0^2 + a(q) + b(q) + \Delta_0^2 \pm 2\sqrt{B_0^2\Delta_0^2 + b(q)(B_0^2 + a(q))}}$$
(S25)

where $a(q) = (2\beta_1 t_1\sin(q) + 2\beta_2 t_2\sin(2q))^2$, $b(q) = (\mu - 2\alpha_1 t_1\cos(q) - 2\alpha_2 t_2\cos(2q))^2$ corresponds to replacing the $2\beta t\sin(q) \to 2\beta_1 t_1\sin(q) + 2\beta_2 t_2\sin(2q)$ and $2\alpha t\cos(q) \to 2t_1\alpha_1\cos(q) + 2t_2\alpha_2\cos(2q)$ in the simple chain problem.

The topological index Pfaffian of the problem also comes out as a simple substitution but since the harmonics of the next nearest neighbor term is $\cos(2q)$, this term does not change sign between $q = 0$ and $q = \pi$. The final condition for nontrivial phase is given by following expression

$$\sqrt{\Delta_0^2 + (\mu + 2\alpha_1 t_1 - 2\alpha_2 t_2)^2} > |B_0| > \sqrt{\Delta_0^2 + (\mu - 2\alpha_1 t_1 - 2\alpha_2 t_2)^2}.$$
(S26)

Note that the existence of a next nearest coupling, in addition to possibly providing favorable magnetic configuration, effectively shifts the chemical potential by $-2\alpha_2 t_2$.



\end{document}